\documentclass[final]{elsarticle}

\usepackage{lineno,hyperref}
\usepackage{tcolorbox}%
 \usepackage{epstopdf}

\journal{Journal of \LaTeX\ Templates}

\begin{document}

\begin{frontmatter}

\title{Optical modes of multilayered photonic structure containing nematic layer with abnormal electroconvective rolls}

\author[1]{Vladimir A. Gunyakov\corref{Corresponding author}} 
\cortext[Corresponding author]{Corresponding author}
\ead{gun@iph.krasn.ru}
\author[1,2]{Mikhail N. Krakhalev}
\author[1,2]{{Ivan V. Timofeev}}
\author[1]{Victor Ya. Zyryanov}
\author[1]{and Vasily F. Shabanov}
\address[1]{Kirensky Institute of Physics, Federal Research Center KSC SB RAS,  Krasnoyarsk, 660036, Russia}
\address[2]{Institute of Engineering Physics and Radio Electronics, Siberian Federal University, Krasnoyarsk, 660041, Russia}

\begin{abstract}
Optical modes of a multilayered photonic structure with the twisted nematic liquid crystal as a defect layer have been investigated. The electroconvective flow in the nematic makes a spatially periodic structure in the form of abnormal rolls. Non-adiabatic propagation of polarized light in the defect layer causes unique features of the optical modes corresponding to the ordinary $o$-waves. The decay of these modes has been demonstrated with increasing voltage due to the effect of cross-polarization diffraction loss. The modes short-wave shift resulting from the contribution of the non-adiabatic geometric phase to the total phase delay of the wave during a round-trip propagation through the photonic structure has been found {both experimentally and numerically}. 
\end{abstract}

\begin{keyword}
\texttt{Photonic structure}
\sep Optical modes
\sep Nematic liquid crystal
\sep Electroconvective instability
\sep Cross-polarization diffraction
\sep Geometric phase

\end{keyword}

\end{frontmatter}


\section{Introduction}

The Fabry-P\'{e}rot cavity-type multilayered photonic structures based on the distributed Bragg mirrors evoke great interest as promising optical materials for nanophotonics and optoelectronics functional elements \cite{joannopoulos_photonic_2008,busch_photonic_2006}. An important property of such structures is the presence of photonic bandgap (PBG) with the low density of photonic states and low transmission. The bandgap exhibits specific dispersion properties, which can be used to implement the regimes of light propagation in photonic structures that are interesting for both fundamental research and application \cite{joannopoulos_photonic_2008,busch_photonic_2006,busch_periodic_2007}. In the multilayered photonic structure with broken periodicity, the narrow resonant peaks called defect (localized) modes appear in the bandgap. Using liquid crystals (LCs) as defect layers, one can form photonic structures with controlled spectral characteristics \cite{busch_periodic_2007,shabanov_optics_2005,arkhipkin_electro-_2011}. A thin twisted nematic layer with the strongly violated Mauguin's waveguide regime \cite{Mauguin_sur_1911} gives rise to the unique spectral features of multilayered photonic structures. The light waves linearly polarized parallel or perpendicular to the optical axis (director \textbf{n}) at the nematic layer input become elliptically polarized when the waveguide propagation regime is violated. The eigenmodes of a twisted nematic are the extraordinary ($te$) and ordinary ($to$) elliptically polarized waves (twist-modes), which adiabatically follow the local director \cite{yeh_optics_1999}. Then, any polarized beam of light can be decomposed into a linear combination of these two twist-modes. 

It was demonstrated using the coupled wave theory that, in the Fabry--P\'{e}rot cavity with a twisted nematic layer, the \textit{te}- and \textit{to}-modes at the same frequency couple with each other upon reflection from mirrors and create the cavity eigenmode \cite{ohtera_analysis_2000}. The state of polarization and, consequently, the cavity eigenmode type (\textit{re} or \textit{ro}) depend on the relative contribution of the twist-modes to each cavity eigenmode. Despite of the ellipticity of the $re$ and $ro$~eigenmodes, their polarizations on the mirrors remain linear in the mutually perpendicular directions \cite{patel_anticrossing_1991,yoda_analysis_1997}. In Ref. \cite{gunyakov_polarization_2017}, the electrically-operated polarization exchange of the cavity eigenmodes in the spectral points corresponding to the Gooch--Tarry maxima \cite{gooch_optical_1975} was demonstrated for a uniformly twisted nematic. With increasing voltage, one can observe series of avoided crossings of the transmission peaks in the spectrum, which cause a bisector (at angles of $\pm45^{\circ}$ to the nematic director) polarization of both cavity modes. The enhancement of the mode coupling leads to the short-wave shift of the \textit{ro}-modes \cite{gunyakov_electric_2018}. Vice versa, in the vicinity of the spectral points corresponding to the Gooch--Tarry minima, the weak mode coupling does not affect the spectral position of the \textit{ro}-modes. Nevertheless, under certain conditions and at the Gooch--Tarry minima, the shift of the \textit{ro}-modes can be observed. In particular, at the transition from the homeoplanar to twist configuration of the director field, anomalous shifts of the \textit{ro}-modes to the short-wave spectral region were reported \cite{timofeev_geometric_2015}. It was established that observed shifts are caused by  contribution of the non-adiabatic geometric phase \cite{aharonov_phase_1987} to the total phase delay of the wave during a round-trip propagation. 

The twist deformation of the director in the spatially periodic structures, e.g. electroconvective nematic LC structure in the form of abnormal rolls, was predicted theoretically \cite{plaut_new_1997} and monitored experimentally \cite{rudroff_relaxation_1999,dennin_direct_2000}. The wave vector \textbf{q} characterizing the periodicity of abnormal rolls is parallel to the orientation of the director \textbf{n} on the substrates as in the case of normal rolls (Williams domains \cite{williams_domains_1963}). While the director projection at the center of abnormal rolls makes an angle with \textbf{n} on the substrates and the regions with different twist signs are separated by domain boundaries. The optical contrast between different domains observed using a circular analyzer is resulted from coupling of the extraordinary (\textit{e}) and ordinary (\textit{o}) waves in the twisted nematics. The wave coupling effect in magnetic field twisted nematic director structures (\textit{T}-effect) leading to the ellipticity of the incident linearly polarized light was described previously by Gerber and Schadt \cite{gerber_measurement_1980}. Under the assumption of smallness of the director gradients, an analytical expression was derived for describing the oscillating behavior of the transverse electrical component $E_{y}$~of the optical eigenmode for LC layer with the anisotropy $\Delta n = n_{e} - n_{o}$ and thickness $d$ as a function of the parameter $w = k_{0}d\Delta n/2\pi$. At the wavelengths $\lambda = 2\pi/k_{0}$ that satisfy the condition $w = 1, 2, 3 .... $ the ellipticity of the modes at the sample output is maximum \cite{gerber_measurement_1980}. 

The diffraction in the electroconvective structures is due to periodic optical inhomogeneities in the~form of phase gratings \cite{carroll_liquidcrystal_1972}. The hierarchy of the increasingly complex convective structures in the LCs easily switched by an electric field and their polarization-sensitive optical response make it possible to control the~mode amplitudes in the transmission spectrum of photonic structures with a nematic defect layer \cite{gunyakov_modulation_2016}. At the same time the question related to the transformation of the optical modes of photonic structure under the coupling effect of waves and their diffraction remains understudied. In view of the aforesaid, here we examine the transformation of the polarized transmission spectra of a multilayered photonic structure with a defect layer of the electroconvective nematic with the abnormal roll instability. A homogeneous twist-deformation (torsion effect) of the director field in the abnormal roll structure is controlled by analyzing the states of polarization of the laser radiation diffracted on the LC cell with the electroconvective nematic layer identical to the one in the photonic structure.

\section{Experimental approach}

\begin{figure} [t!]
\centering
\includegraphics[width=1.0\linewidth]{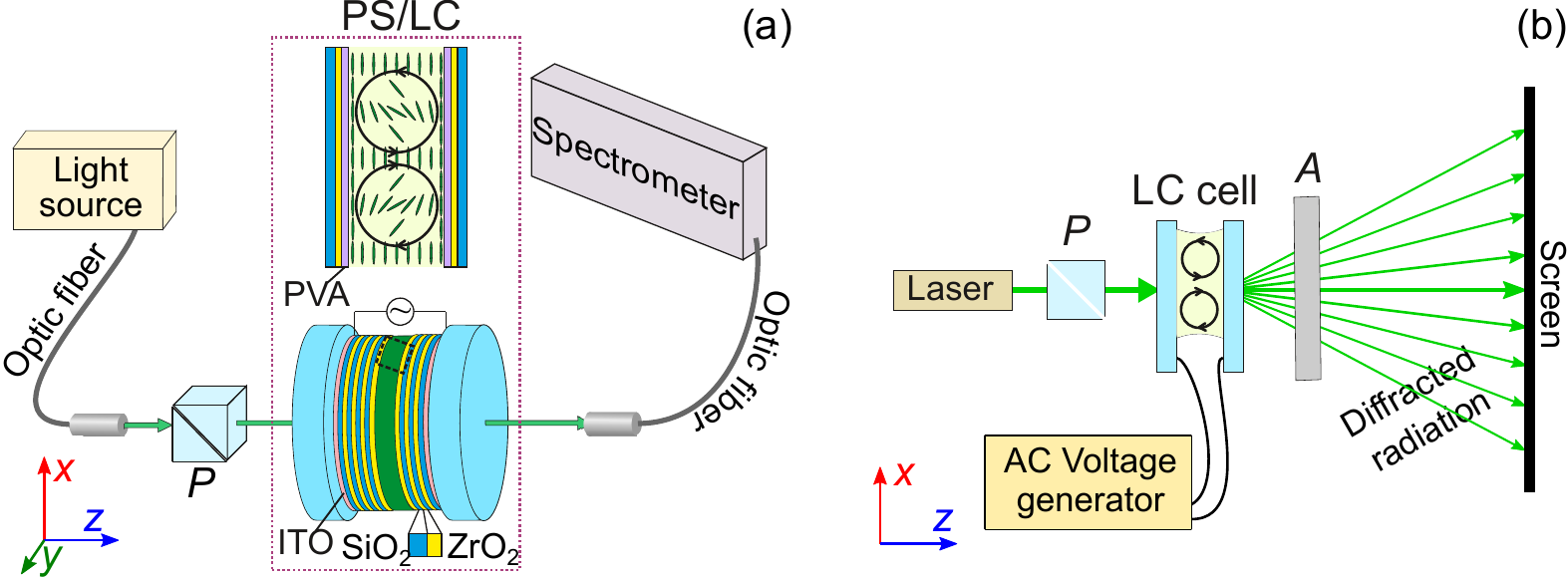}
\caption{Experimental setups for recording the polarized transmission spectra of the photonic structure with liquid crystal (PS/LC) (a) and diffraction pattern from the LC cell (b) with the convective rolls in nematic layer. Vortex flow of the nematic is indicated by circular arrows.}
\label{fig1}
\end{figure}

\begin{figure} [t!]
\centering
\includegraphics[width=0.50\linewidth]{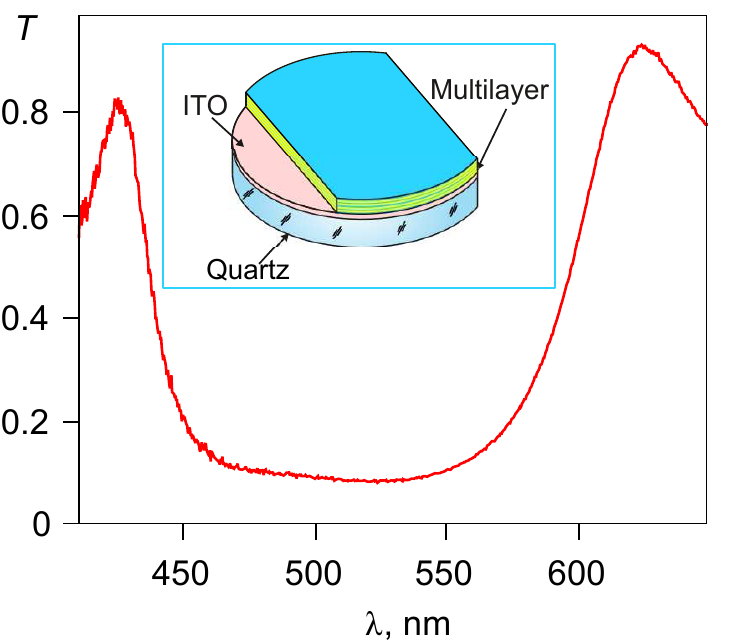}
\caption{{Transmission coefficient of the single Bragg mirror as a function of the light wavelength under normal incidence. Scheme of the mirror is shown on insert.}}
\label{fig2}
\end{figure}

{Figure~\ref{fig1} shows two schemes of experimental setups for studying the spectral characteristics of the directly transmitted radiation and the polarization parameters of a diffracted laser beam. Transmission spectra of the photonic structure were measured by setup presented in Fig.~\ref{fig1}a. The photonic structure with LC defect layer (PS/LC) is composed of two identical Bragg mirrors. Bragg mirrors were made in Technological Design Institute of Applied Microelectronics (Novosibirsk) by electron-beam evaporation technique. Six ZrO$_2$ layers and five SiO$_2$ layers were deposited alternately. It should be noted that available oxide dielectric materials, e.g. TiO$_2$/SiO$_2$ pair has higher refractive indexes difference \cite{CHIASERA2019107} and, respectively, higher quality factor \textit{Q} of the PS/LC than ZrO$_2$/SiO$_2$ \cite{Jerman:05}. The fact is that in real photonic structures an increase of the \textit{Q}-factor leads to the decrease of resonant transmission peaks due to scattering and absorption inside the cavity. And strongest decay of these modes occurs at the center of the PBG \cite{Arkhipkin2008}. That is why we use ZrO$_2$/SiO$_2$ mirrors. This circumstance may be crucial for the determination of the spectral properties of optical modes in photonic structures with electroconvective nematics that are highly scattering media. 
Transmission spectrum and a schematic view of the Bragg mirror are presented in Fig.~\ref{fig2}. 
The multilayer is deposited onto fused quartz substrate with ITO-electrode. The thickness of each layer measured by TEM microscopy on the Bragg mirror was $63~\pm~5$~nm and $82~\pm~5$~nm for the ZrO$_2$ and SiO$_2$, respectively, and a thickness of $168~\pm~5$~nm for the ITO layer (see Supplementary). The first order bandgap ranges from 424~nm to 624~nm. The minimal transmission is about 0.08 and smoothly increases toward the PBG edges. The observed transmittance difference of the short- and long-wave PBG edges is related to optical properties of the ITO thin film that starts to loss transparency for wavelengths shorter than 450~nm \cite{Krylov2013}.}

A gap between the mirrors was filled with a 4-methoxybenzylidene-4'-butyl aniline (MBBA) nematic LC with the negative permittivity anisotropy ($\epsilon_a < 0$) and positive conductivity anisotropy ($\sigma_a > 0$). 
The gap thickness was $7.8~\pm~0.1~\mu$m, {operating area of the PS/LC was $16~\times~22$~mm$^2$}. The nematic clearing point is $T_c = 45^\circ$. 
The Bragg mirrors were coated with rubbed polyvinyl alcohol (PVA) films to specify the planar alignment of the MBBA director \textbf{n}~$\parallel x$. The uniform alignment of the director ensured the transparent state of the LC under the low-frequency (100 Hz) voltage in the range of 0~$\div$~7.2~V. With that the spectral properties of the photonic structure remain unchanged. A critical voltage of $U_c = 7.2$~V leads to the convective instabilities in the form of the spatially periodic flows of the MBBA, which significantly change the optical properties of the LC and thereby modify the optical response of the photonic structure.

Let an incident beam of light be polarized parallel (perpendicular) to \textbf{n} at entrance plane of the twisted nematic layer. This is known as the \textit{e}-mode (\textit{o}-mode) operation \cite{yeh_optics_1999}. The respective polarized components $T_{e,o}(\lambda)$ of the directed transmission spectra and their field-effect evolution were studied using an Ocean Optics HR4000CG spectrometer equipped with fiber optics (Fig.\,\ref{fig1}a). A collimator of the receiving light guide was used as a diaphragm with an input aperture diameter of 4~mm. A Glan prism (\textit{P}) was used as a polarizing element. Spectra were recorded at a fixed temperature of $25~^\circ$C; the sample thermal stabilization accuracy was not worse than $\pm 0.2~^\circ$C. 
	
To elucidate the effect of diffraction loss on the spectral characteristics of the investigated photonic structure, the electroconvective LC cell with two glass substrates coated with ITO electrodes and a nematic layer identical to the one in photonic structure was assembled. Figure\,\ref{fig1}b shows a scheme of the experimental setup used to observe the diffraction pattern that occurs when the polarized laser beam passes through an electric field-controlled LC cell. We used a Newport R-30972 single mode He-Ne laser with an operation wavelength of $\lambda = 543$~nm. The states of polarization of the diffracted radiation were analyzed using a large-format polaroid (\textit{A}), covering all the observed diffraction orders.

\section{Results and Discussion}

\begin{figure} [t!]
\centering
\includegraphics[width=1.0\linewidth]{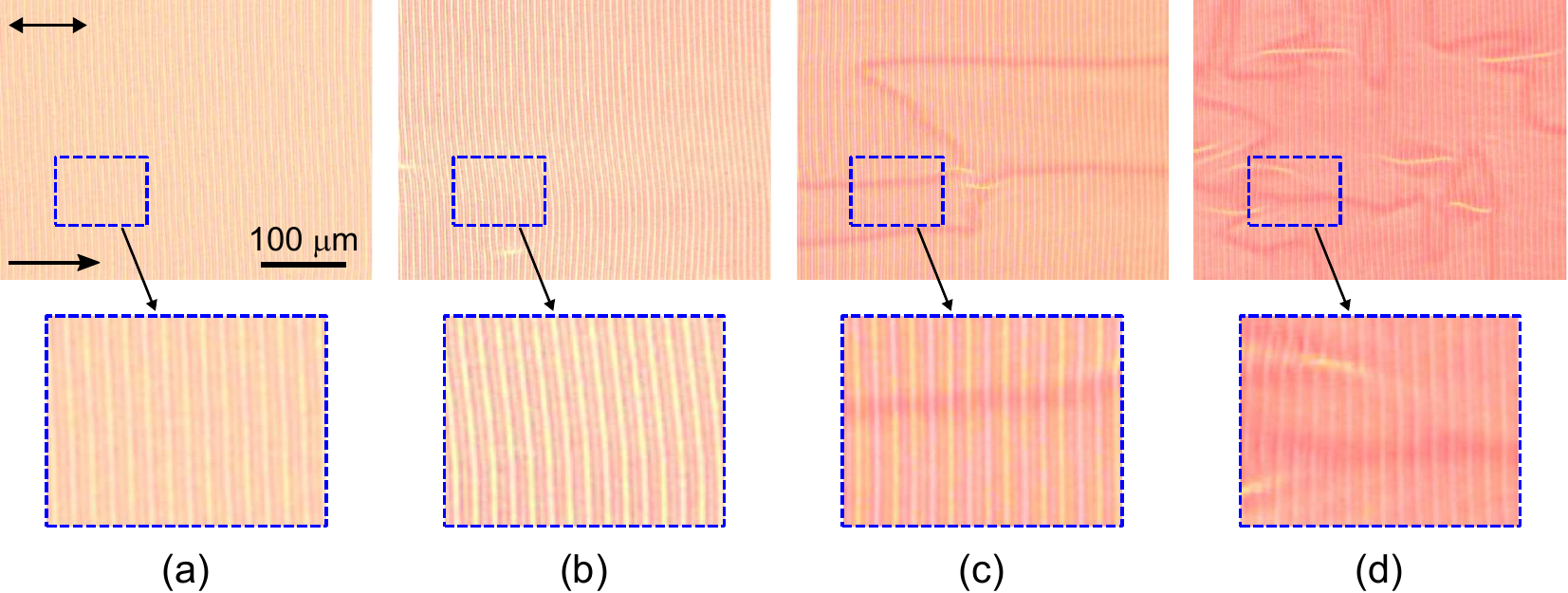}
\caption{Snapshots of the photonic structure with convective rolls at applied ac voltages of 7.3~V (a), 7.6~V (b), 8.0~V (c), and 10.0~V (d). Double arrow ($\leftrightarrow$) indicates the polarizer direction, single arrow ($\rightarrow$) indicates the rubbing direction.}
\label{fig3}
\end{figure}

\begin{figure} [t!]
\centering
\includegraphics[width=1.0\linewidth]{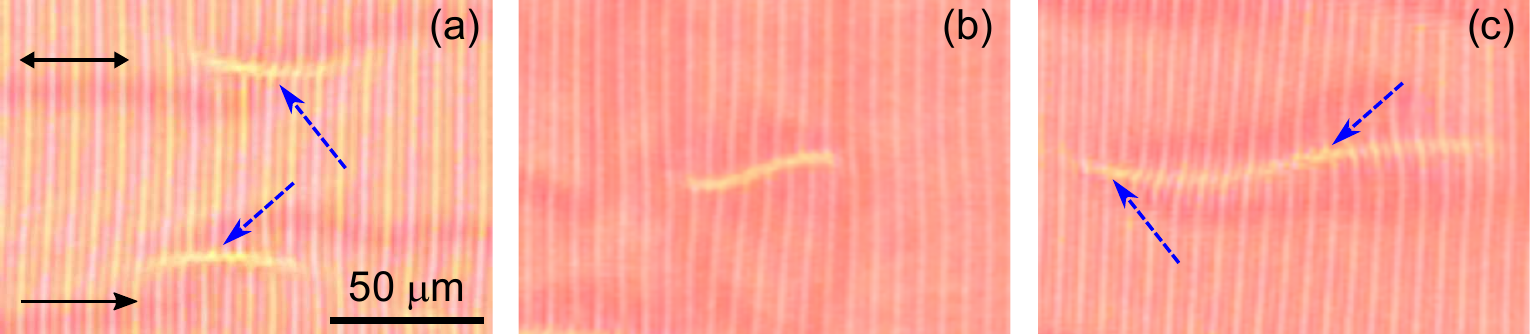}
\caption{{Snapshots of the rolls structure defects at applied ac voltages of 8.0~V (a), 10.0~V (b) and 10.0~V (c).
The pair of opposite edge dislocations (a), screw varicose line (b), combination of the pair of opposite edge dislocations and screw varicose line (c).
Dashed single arrows indicate edge defects, double arrow indicates the polarizer direction, solid single arrow indicates the rubbing direction.
}}
\label{fig4}
\end{figure}

Experimental snapshots of the photonic structure with nematic obtained using a polarizing microscope show the evolution of the electroconvective instability in LC with increasing voltage (Fig.\,\ref{fig3}a-d). The highest contrast of convective rolls is reached when the polarizer is parallel to the rubbing direction, i.e. for the $e$-mode operation. Immediately above the critical voltage $U_c$ in the range of 7.2--7.4~V the normal rolls (Williams domains) are observed in the MBBA layer (Fig.\,\ref{fig3}a). As the applied voltage increases, a weak undulating disturbance of normal rolls arises in the sample (Fig.\,\ref{fig3}b). Above a voltage of 7.6~V, the rolls gradually straighten and transform into abnormal ones. Simultaneously, domain boundaries are formed, which now cross straightened abnormal rolls (Fig.\,\ref{fig3}c).
{The domain boundaries separating regions of stationary rolls with different twist signs loop close on roll structure defects}.
{Two types of such defects are observed: topologically stable edge dislocations Fig.~\ref{fig4}a, non-topological linear defects -- a~localized screw varicose Fig.~\ref{fig4}b, as well as their combination Fig.~\ref{fig4}.}
At the same time both the {roll structure} defects and domain boundaries are characterized by the chaotic dynamics at a fixed voltage \cite{krekhov_spatiotemporal_2015}.
{Similar to the normal roll structure defects \cite{Joets1991}, the opposite dislocation can annihilate. At the same time, a non-topological screw varicose may appear. This linear defect either generates a pair of edge dislocations or disappears (see, for example, Supplementary). At a fixed voltage, the length and density of defects in the structure of the rolls remain on average constant.}
As the applied voltage increases, the density of {the roll structure} defects and their travel rate significantly increase (Fig.\,\ref{fig3}d).

\begin{figure} [t!]
\centering
\includegraphics[width=0.60\linewidth]{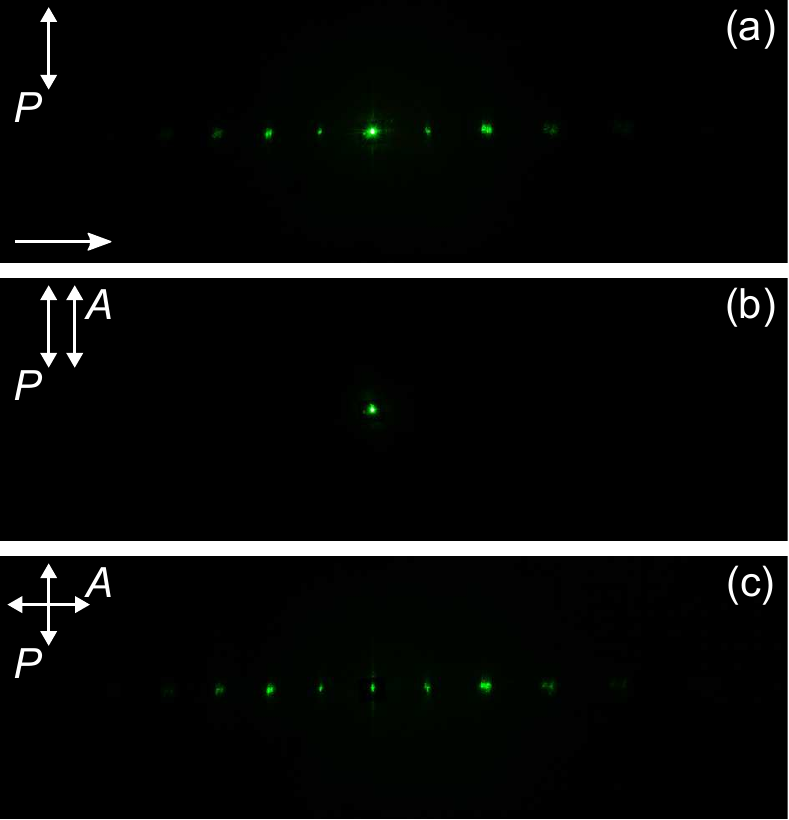}
\caption{Snapshots of far-field diffraction patterns from the LC cell with abnormal rolls for the $o$-mode operation without analyzer (a), with analyzer oriented parallel (b) and perpendicular (c) to polarizer. The applied voltage was $U = 8.0$~V.}
\label{fig5}
\end{figure}

A widespread approach to the direct observations of the twist-deformation of the nematic director field in the electroconvective structures like abnormal rolls is based on a polarization-optical technique using a circular analyzer \cite{rudroff_relaxation_1999,dennin_direct_2000}. In addition, it is convenient to analyze the states of polarization of the laser radiation diffracted on the roll structure in LC layer. In our case, this method is more relevant, since the decay of the $re$- and $ro$-modes is obviously related to the diffraction loss of light
propagating through the roll structure. In particular, under illumination of an LC cell with convective twist domains (Fig.\,\ref{fig3}c) by a polarized laser beam, a pattern of diffraction reflexes is observed in the plane perpendicular to the roll axis (Fig.\,\ref{fig1}b). Moreover, the diffraction arises not only in the trivial case of the $e$-mode operation \cite{carroll_liquidcrystal_1972}, but also at the laser beam polarization parallel to the axis of abnormal rolls, i.e. for the $o$-mode operation. Indeed, in the last case $N^{th}$-order diffraction reflexes ($N = \pm 1, \pm 2, \pm 3, \ldots$) are observed on the left and on the right of directly transmitted beam (Fig.\,\ref{fig5}a). A large-format analyzer oriented parallel to the polarizer decreases slightly the brightness of the zero-order reflex, while the higher-order reflexes become invisible (Fig.\,\ref{fig5}b). Vice versa, when analyzer is oriented perpendicular to the polarizer the brightness of the higher-order reflexes is practically unchanged, while the zero-order reflex brightness decreases significantly (Fig.\,\ref{fig5}c). It means that the deflected beams generating all higher-order reflexes are polarized perpendicular to the polarizer direction (cross-polarization diffraction). Obviously, only twisting the director \textbf{n} in the LC layer can lead to the diffraction for the $o$-mode operation and to the polarization change of diffracted light in such a way. It should be recalled, in the case of normal rolls the beams of light produce no deflects for the $o$-mode operation \cite{carroll_liquidcrystal_1972}. Note the stability of the observed diffraction pattern, in spite of the convective instability which is characterized by the active dynamics of {the roll structure} defects and domain boundaries (see Fig.\,\ref{fig3}c). A large number of the twist domains is located in the cross-section of a laser beam with a ($1/e^2$) diameter of 0.83~mm. Nevertheless, the periodicity of abnormal rolls $\Lambda = 7.4~\mu$m  is preserved at a fixed voltage $U = 8.0$~V; therefore, the angular distribution of the diffraction maxima remains constant.

\begin{figure} [t!]
\centering
\includegraphics[width=1.0\linewidth]{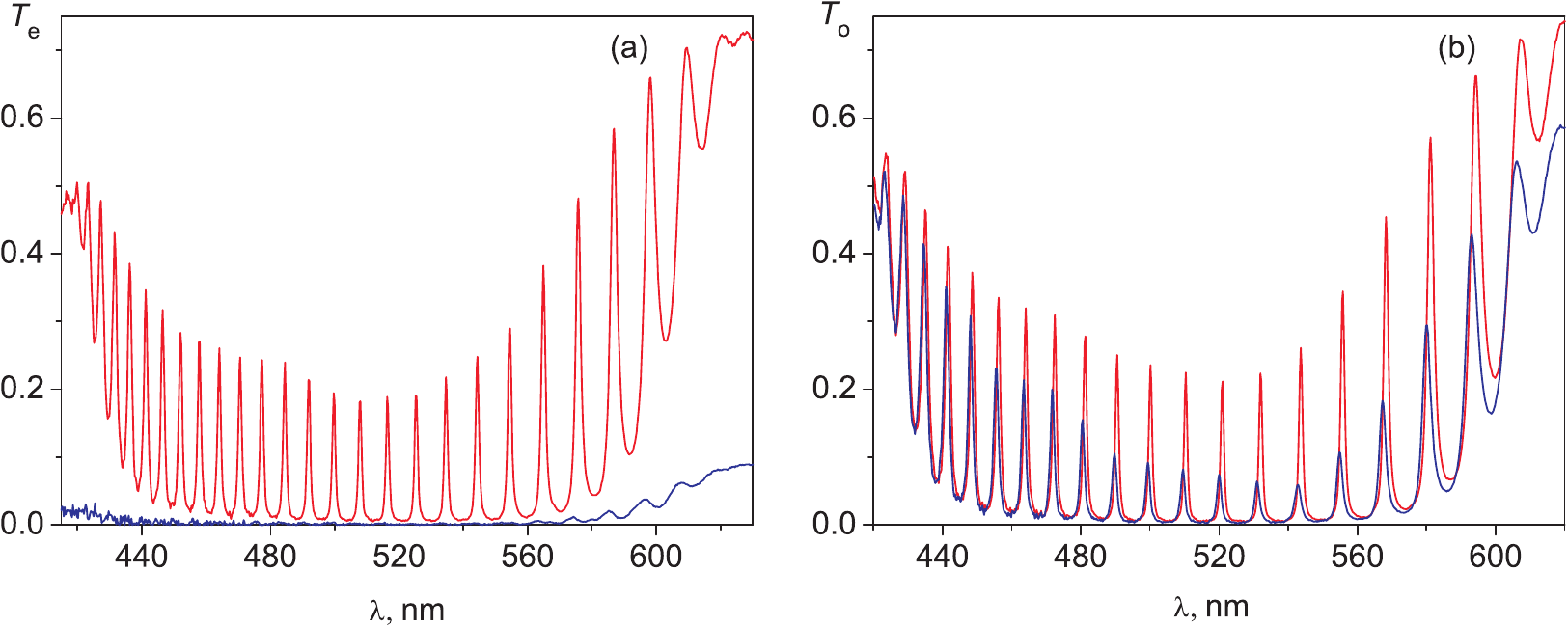}
\caption{Polarized components of the directed transmission spectra $T_{e,o}(\lambda)$ of the multilayered photonic structure for the $e$-mode (a) and $o$-mode (b) operation. Red lines correspond to the voltage range of $U \leq U_c$, blue lines correspond to the voltage $U = 10$~V.}
\label{fig6}
\end{figure}

The polarized components of the directed transmission spectra $T_{e,o}(\lambda)$ of the photonic structure at the voltages $U < U_c$ and $U = 10$~V are shown in Figure\,\ref{fig6}. Due to the periodicity of the photonic structure the bandgap is formed with a set of localized modes. Their spectral positions are determined by optical properties of the LC defect. Fig.\,\ref{fig6} shows different response of the $re$- and $ro$-modes of the photonic structure respond differently to the electroconvection process occurring in the nematic layer. By the time when the voltage increases to 10 V, the amplitudes of the $re$-modes damp to the PBG background level, while the signal level at the edges of the bandgap itself decreases by an order of magnitude (Fig.\,\ref{fig6}a). At voltages $U > U_c$ the extraordinary waves diffract on the 1D phase grating of normal and abnormal rolls; therefore, the directed transmission of light decreases due to the spatial redistribution of the wave energy. Thus, the diffraction losses are the main reason for the rapid decay of the $re$-modes \cite{gunyakov_modulation_2016}, starting from a critical voltage of $U_c = 7.2$~V. With voltage increasing, the additional loss comes from {the roll structure} defects and domain boundaries. In particular, the density of {the roll structure} defects increases from 4 defects per~mm$^2$ (Fig.\,\ref{fig3}a) to 35 defects per~mm$^2$ (Fig.\,\ref{fig3}d), while their length increases by an order of magnitude on average. Despite these factors, for the $o$-mode operation, at a voltage of 10 V the bandgap is maintained, and the $ro$-modes are still observed (Fig.\,\ref{fig6}b). The amplitudes of these modes decrease by a factor of 2 on average in the long-wave region and by a factor of 3 at the center of the PBG (Fig.\,\ref{fig6}b). It seems that the decay of the $ro$-modes is related to the loss caused by both scattering on {the roll structure} defects and the cross-polarized diffraction on abnormal rolls in nematic defect of the photonic structure.

\begin{figure} [t!]
\centering
\includegraphics[width=1.0\linewidth]{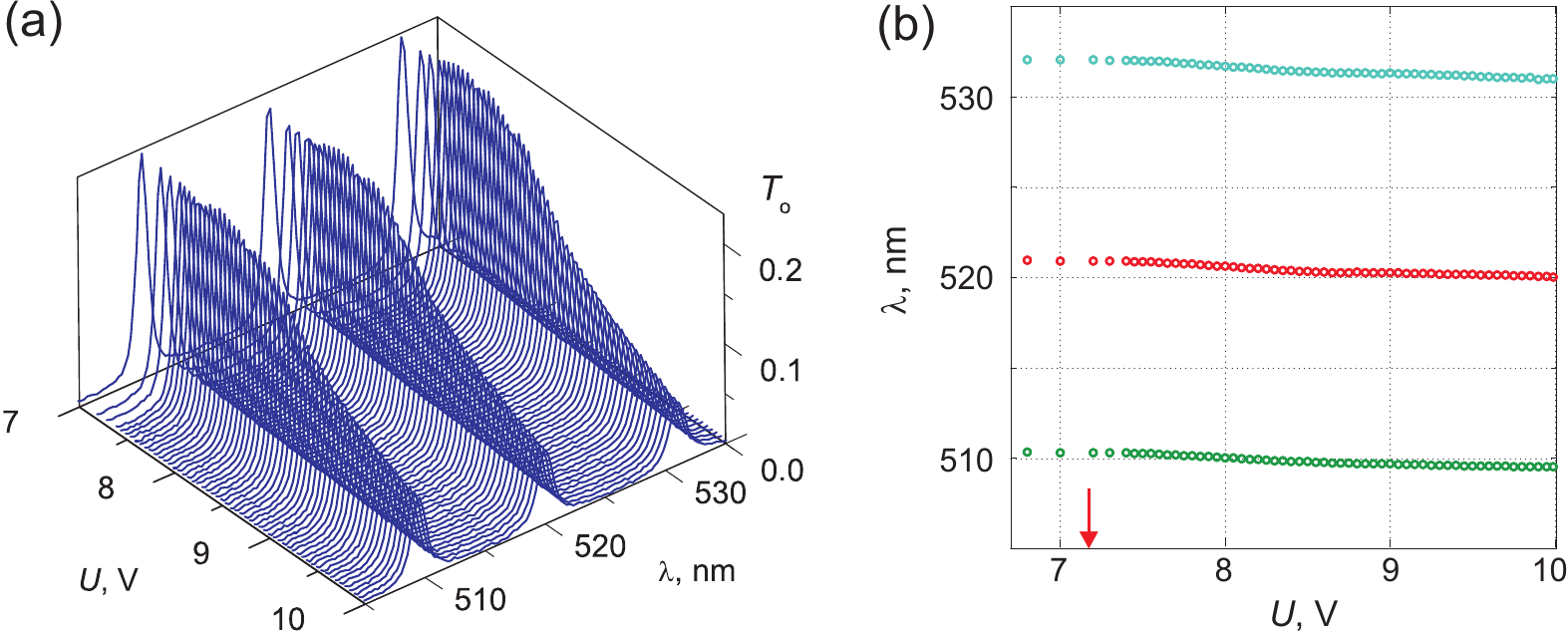}
\caption{Field-effect dependencies of the transmission spectrum of the photonic structure $T_o(U)$ (a) and spectral positions of the $ro$-modes maxima $\lambda(U)$ (b) at the center of the PBG. The arrow indicates the critical voltage $U_c = 7.2$~V.}
\label{fig7}
\end{figure}

Along with a significant decrease in the amplitude, another feature in the behavior of the $ro$-modes at the voltage increasing is their anomalous spectral shift to the short-wave region relative to the initial position (Fig.\,\ref{fig6}b). Field-effect dependencies of the transmittance $T_o(U)$ and spectral positions of the maxima $\lambda(U)$ of the $ro$-modes in the most sensitive central PBG region are shown in Fig.\,\ref{fig7} for the voltage range of $6.8 \leq U \leq 10.0$~V. The measurements were performed with a step of 0.05~V at a fixed temperature. Up to the voltage value of 7.8~V, the amplitudes and spectral positions of the $ro$-mode peaks remain insensitive to the convective instabilities and have the same values as for the initial state of nematic. Above a voltage of 7.8~V, the abnormal rolls are observed. They are characterized by the presence of a homogeneous twist deformation of the director \textbf{n}. As the applied voltage increases, the gradually increased twist angle affects the decay of the $ro$-mode amplitudes (Fig.\,\ref{fig7}a) and their smooth spectral shift (Fig.\,\ref{fig7}b). Obviously, the synchronization of both effects is related to the occurrence of the elliptically polarized $te$- and $to$-modes, which redistribute the light fluxes in the photonic structure. The asymptotic behavior of the $\lambda(U)$ curves in Fig.\,\ref{fig7}b evidences for  stabilization of the twist-angle value upon approaching a point of $U = 10$~V. Further decay of the $ro$-modes at voltages above 10~V is mainly caused by increasing the density and length of the {roll structure} defects (Fig.\,\ref{fig3}d). The maximal shift of the $ro$-modes to the short-wave region amounts to 0.9~nm on average. For the twisted nematic LC layer the probing radiation at some wavelengths has a zero transverse component $E_y~(E_x)$ of the optical eigenmode at the output of the sample \cite{gerber_measurement_1980}. This means that the elliptical polarization of light inside the LC layer transforms to linear and parallel (perpendicular) to the output director \textbf{n}. The $re$- and $ro$-modes of the photonic structure at these wavelengths are polarized in a similar way. At that the effect of coupling the twist-modes upon reflection from mirrors is not revealed. It should be noted that $ro$-modes shown in Fig.\,\ref{fig7}b remain linearly polarized perpendicular to the director \textbf{n} on the mirrors. Therefore, the observed anomalous shift of the $ro$-modes to the short-wave region is most likely the spectral manifestation of the non-adiabatic geometric phase. After the voltage switched-off the director field configuration returns to the initial state. In this case, both the amplitudes and spectral positions of the $ro$-modes are recovered.

{The directed transmission spectrum of the \textit{ro}-modes in the multilayered photonic structure has been simulated within one-dimensional approximation of the homogeneous director twist deformation in abnormal rolls \cite{krekhov_spatiotemporal_2015,rudroff_relaxation_1999}:}

\begin{equation}
{\phi = \phi_0 sin(\pi z/d),~~\theta = \theta_0 sin(\pi z/d).}
\end{equation}

\begin{figure} [t!]
\centering
\includegraphics[width=0.50\linewidth]{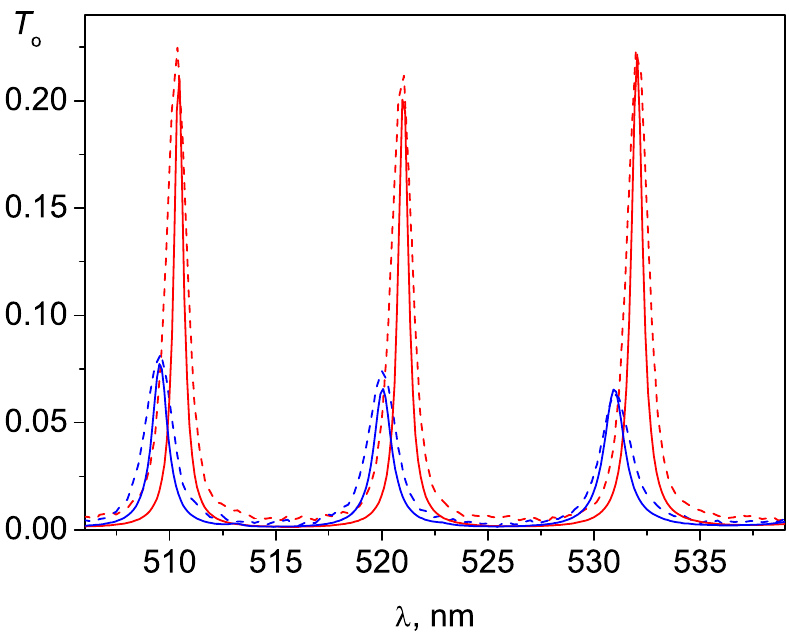}
\caption{{Spectral position of the modes at the  PBG center without (red lines) and under (blue lines) voltage for \textit{o}-mode operation, measured (dashed lines) and simulated using the 4$\times$4 transfer matrix method (solid lines).}}
\label{fig8}
\end{figure}

{Here $\phi_0$, $\theta_0$ are maximal azimuthal and polar angles of the director in the midplane of a layer, respectively: $0~\leq~z~\leq~d$. 
For $U=0$ we take $\theta_0=\phi_0=0$, and for $U=10$~V we assume $\theta_0=24^\circ,  \phi_0=60^\circ$ \cite{krekhov_spatiotemporal_2015}.
Then, using the 4$\times$4 transfer matrix method \cite{Berreman:72},
the transmission spectrum for \textit{o}-mode operation in the investigated multilayer structure is simulated with regard to the optical extinction and material dispersion. 
The following parameters are used. Each of two mirrors is a stack of the SiO$_{2}$ substrate, ITO electrode, ZrO$_{2}$, (SiO$_{2}$, ZrO$_{2})^{5}$, and PVA layer. The thicknesses and refractive indices of the amorphous layers of the dielectric mirrors are 80~nm and 1.45 for SiO$_{2}$, 65~nm and 2.04 for ZrO$_{2}$, 1.515 and 20~nm for the PVA layer. The values for the ITO layer are 160~nm and 1.88858+0.022i with account of the extinction; the substrate refractive index is 1.45 and MBBA  refractive indices are $n_{||}$~=~1.765 and $n_{\perp}$~=~1.553, respectively (the wavelength is $\lambda $~=~500~nm and the temperature is $T = 25^\circ$~C). The nematic layer thickness is 7522~nm. Taking into account the dispersion, we use the data and reference from https://refractiveindex.info.
The damp of transmission peaks is approximated by increased LC extinction. 
The imaginary part of MBBA refractive index is taken 4$\cdot $10$^{-4}$i for $U=0$ and 8.5$\cdot $10$^{-4}$i for $U=10$~V.}
{ Experimental and calculated transmission spectra are presented in Figure~\ref{fig8} . 
It can be seen that the experimental and calculated spectral positions of the resonator modes are in good agreement for the $\phi_0$ and $\theta_0$ angles used in simulation.}

\section{Conclusion}

Thus, we experimentally investigated the transformation of optical modes of the multilayered photonic structure with a defect layer of the electroconvective nematic LC with the abnormal roll instability. In contrast to the well-known Williams domains, this instability is characterized by the homogeneous twist deformation of the director field \cite{plaut_new_1997}. This leads to the ellipticity of the linearly polarized light travelling through the disturbed nematic. The twist deformation is identified by analyzing the polarization properties of the laser radiation diffracted on the electroconvective LC cell with a nematic layer. In particular, the diffraction for the $o$-mode operation was established. In this case the diffraction pattern is a superposition of zero-order reflex corresponding to the wave linearly polarized perpendicular to the director and zero- and higher-order reflexes corresponding to the wave polarized along the director (cross-polarized diffraction). Under the voltage growth the cross-polarized diffraction results in smooth decay of the $ro$-modes. The anomalous shift of the $ro$-modes to the short-wave range was found, which can be attributed to the contribution of the non-adiabatic geometric phase to the total phase delay experienced by the wave during a round-trip propagation. 
{The numerical simulation of transmission spectrum of structure with homogeneous twist deformation of the director that is typical for abnormal rolls demonstrates the same effect.} 
The proposed approach based on using a multilayered photonic structure can be used for the study of the features of spatially periodic structures in dissipative liquid-crystalline systems with a complex director field configuration because of  high sensitivity of the optical modes to electroconvective processes. This optical material is promising for the applications in various optoelectronic devices: tunable spectral filters, polarizing sensors, etc.

\section*{Acknowledgments}
{We are grateful to M.N. Volochaev for providing TEM micrograph of the mirror.}


\bibliography{References}

\end{document}